%% file: CloneRefactoring.tex
\documentclass[conference]{IEEEtran}
\usepackage{flushend}
\usepackage{booktabs}
\usepackage{comment}
\usepackage[hyphens]{url}
\usepackage{graphicx}
\usepackage{xspace}
\usepackage[dvipsnames]{xcolor}
\usepackage{color}
\usepackage{listings}
\usepackage{multirow}
\usepackage{array}
\usepackage{tcolorbox}
\usepackage{amssymb,amsmath,amsthm}
\usepackage{marvosym}
\usepackage{balance}
\usepackage{MnSymbol}
\usepackage{subfig} 
\usepackage{pifont}
\usepackage[flushleft]{threeparttable}
\usepackage{cite}

\input{macros}

\newcommand{\tool}{\textsc{CRec}\xspace}

\newcommand{\codefont}[1]{\footnotesize{\texttt{#1}}\normalsize}
\newcommand{\fcount}{34}
\newcommand{\rclone}{R-clones}
\newcommand{\nrclone}{NR-clones}

\newcolumntype{R}[1]{>{\raggedleft\let\newline\\\arraybackslash\hspace{0pt}}m{#1}}
\ifCLASSINFOpdf
\else
\fi
\graphicspath{{images/}}

\begin{document}
%
\title{Automatic Clone Recommendation for Refactoring Based on the Present and the Past}

\author{\IEEEauthorblockN{
		Ruru Yue\IEEEauthorrefmark{1}$^{,}$\IEEEauthorrefmark{2}\hspace*{1.5em}
		Zhe Gao\IEEEauthorrefmark{1}$^{,}$\IEEEauthorrefmark{2}\hspace*{1.5em}
		Na Meng\IEEEauthorrefmark{3}\hspace*{1.5em}
		Yingfei Xiong\IEEEauthorrefmark{1}$^{,}$\IEEEauthorrefmark{2}\hspace*{1.5em}
		Xiaoyin Wang\IEEEauthorrefmark{4}\hspace*{1.5em}
		J. David Morgenthaler\IEEEauthorrefmark{5}
}
	\IEEEauthorblockA{
		\IEEEauthorrefmark{1}Key Laboratory of High Confidence Software Technologies (Peking University), MoE
		\\ \IEEEauthorrefmark{2}Institute of Software, EECS, Peking University, Beijing 100871, China
		\\\{yueruru,sonia,xiongyf\}@pku.edu.cn
		\\ \IEEEauthorrefmark{3}Department of Computer Science, Virginia Tech, Blacksburg, VA, 24060, nm8247@cs.vt.edu
		\\ \IEEEauthorrefmark{4}Department of Computer Science, University of Texas at San Antonio, San Antonio, TX 78249, xiaoyin.wang@utsa.edu
		\\ \IEEEauthorrefmark{5}Google, Mountain View, CA 94043, jdm@google.com
	}
}

%


\maketitle

\input{abstract}


%
\IEEEpeerreviewmaketitle

\input{intro2}


\input{approach}

\input{results}

\input{related_work}

\input{threats}

\input{conclusion}

\section*{Acknowledgment}
We thank anonymous reviewers for their thorough and valuable comments on our earlier version of the paper. This work is sponsored by the National Key Research and Development Program under Grant No.~2017YFB1001803, the National Natural Science Foundation of China under Grant No.~61672045 and 6133201, NSF Grant No.~CCF-1565827, ONR Grant No.~N00014-17-1-2498, NSF Grant No.~CNS-1748109 and DHS Grant No.~DHS-14-ST-062-001.
\bibliographystyle{abbrv}
\bibliography{refactoring}

\end{document}

%% file: macros.tex
\usepackage{listings}

\newtoggle{longversion}
\toggletrue{longversion}

\newtoggle{patchtype}
\togglefalse{patchtype}

%% file: abstract.tex
\begin{abstract}

When many clones are detected in software programs, not all 
clones are equally important to developers. 
To help developers refactor code and improve software quality, 
various tools were built to recommend clone-removal refactorings based
on the past and the present information, such as the cohesion degree of individual clones or the co-evolution relations of clone peers. 
The existence of these tools inspired us to build an approach that considers as many factors as possible to more accurately recommend clones.
This paper introduces \tool, a learning-based approach 
that recommends clones by extracting features from the current status and past history of software projects. 
Given a set of software repositories, \tool first automatically extracts the clone groups historically refactored (\rclone) and those not refactored (\nrclone) to construct the training set. 
\tool extracts \fcount{} features to characterize the content and evolution behaviors of individual clones, as well as the spatial, syntactical, and co-change relations of clone peers. 
With these features, \tool trains a classifier that recommends clones for refactoring.   

We designed the largest feature set  thus far for clone recommendation,
and performed an evaluation on six large projects. The results show that
our approach suggested refactorings with 83\% and 76\%
F-scores in the within-project and cross-project settings. \tool significantly outperforms
a state-of-the-art similar approach on our data set,
with the latter one achieving 70\% and 50\% F-scores. We also compared the
effectiveness of different factors and different learning algorithms.



\end{abstract}


%% file: intro2.tex
\section{Introduction}
Code clones (or duplicated code) present challenges to software maintenance. They may require developers to repetitively apply similar edits to multiple program locations.
When developers fail to consistently update clones, they may incompletely fix a bug or introduce new bugs~\cite{Li2004:cpminer, jiang2007context,Park2012:supplementary,Ray2013}. To mitigate the problem, programmers sometimes apply clone removal refactorings (e.g., \emph{Extract Method} and \emph{Form Template Method}~\cite{Fowler2000}) 
to reduce code duplication and eliminate repetitive coding effort.
However, as studied by existing work~\cite{kim2005empirical,gode2011frequency,wang2014recommending}, not all clones need to be refactored. 
Many clones are not worthwhile to be refactored or difficult to be refactored~\cite{kim2005empirical}, and programmers often only refactor a small portion of all clones~\cite{wang2014recommending}.
Thus, when a clone detection tool reports lots of clones in a program, it would be difficult for the developer to go through the (usually long) list and locate the clones that should be refactored. A technique that automatically recommends only the important clones for refactoring would be valuable.

Researchers have realized this problem and various tools were developed to recommend clones for refactoring. Given a project, most current tools recommend clones based on factors either representing \emph{the present status or past evolution of the project}.
For example, some tools suggest refactorings based on the code smell presented in a program, e.g., high cohesion and low coupling of clones~\cite{higo2008metric,Goto2013:rank}.
Some other tools provide recommendation based on the past co-evolution relationship between clones~\cite{Higo2013,hua2015does}.
The
success of these tools indicates that both the present and the past of programs 
are useful in recommending clones, and it naturally raises a question:
can we build an approach that considers as many factors as possible to
more accurately recommend clones?

To take multiple factors into account, we need a model
that aggregates all factors properly. 
Machine learning provides a good way to build a such model.
Existing work~\cite{wang2014recommending}
showed that using machine learning to integrate factors could lead
to good accuracy in recommending clones. However, machine learning
requires a large set of labelled instances for training.
In practice,
we often lack enough human power to manually build such a training set; neither have we found any existing work that automatically builds the set.


\begin{figure*}[!htb]
	\centering
	\includegraphics[width=15cm]{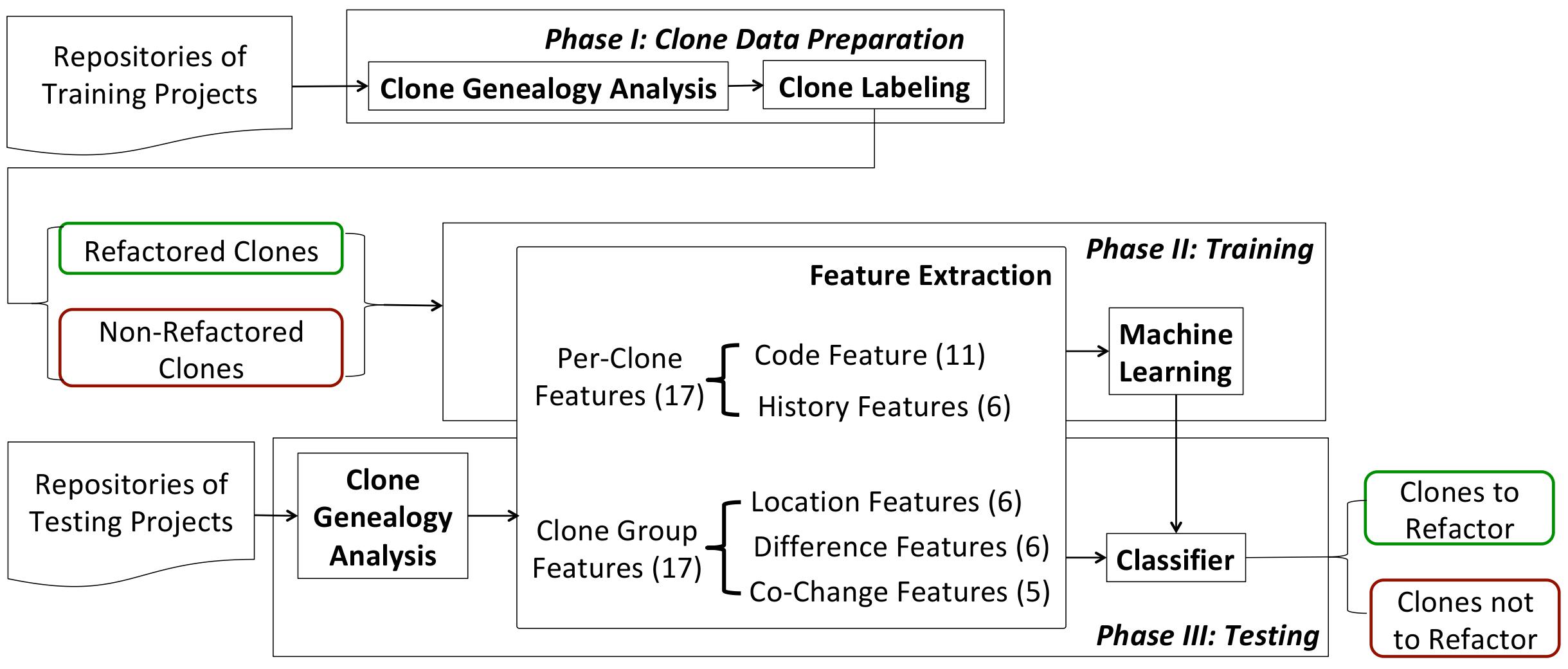}
	\vspace{-0.5em}
	\caption{\tool{} has three phases: clone data preparation, training, and testing. Phase I detects refactored and non-refactored clones in the repositories of training projects.
	Phase II trains a classifier with machine learning based on the two types of clones. Phase III recommends clones for refactoring using the trained classifier.}
	\label{fig:overview}
	\vspace{-1.5em}
\end{figure*}

In this paper, we present \tool{}, a learning-based approach that
automatically extracts refactored and non-refactored clones groups
from software repositories, and trains models to
recommend clones for refactoring. 
Specifically
as shown in Fig.~\ref{fig:overview}, there are three phases in our
approach. Given open source project repositories, \tool first models
clone genealogies for each repository as prior work
does~\cite{kim2005empirical} and identifies clone groups historically
refactored (\textbf{R-clones}) and others not (\textbf{NR-clones}) in
Phase I. 

\tool then extracts \fcount{} features to characterize both kinds of groups and to train a model in Phase II. 
Intuitively, developers may decide refactorings based on (1) the cost of conducting refactorings (e.g., the divergence degree among clones), and (2) any potential benefit (e.g., better software maintenability). 
To consider as many factors as possible, we thus include 17 per-clone
features to characterize each clone's content and evolution history,
and 17 clone-group features to capture the relative location,
syntactic difference, and co-change relationship between clone peers.
The \fcount{} features in total describe both the present status and
past evolution of single clones and clone peers. 

Finally in Phase III, 
to recommend clones given a project's repository, \tool performs clone genealogy analysis, extracts features from the clone genealogies, and uses the trained model for refactoring-decision prediction.


We evaluated \tool{} with R-clones and NR-clones extracted from six open source projects in the
within-project and cross-project settings. Based on our dataset and
extracted features, we also evaluated the effectiveness of different feature subsets and machine
learning algorithms for clone refactoring recommendation. Our
evaluation has several findings. (1)
\tool{} effectively
suggested clones for refactoring with 83\% F-score in the
within-project setting and 76\% F-score in the cross-project setting.
(2) \tool significantly outperforms a state-of-the-art learning-based
approach~\cite{wang2014recommending}, which has 70\% F-score in the within-project setting and
50\% F-score in the cross-project setting. (3) Our automatically
extracted refactored clones achieve a precision of 92\% on average. (4) Clone history features and co-change features are the
most effective features, while tree-based machine learning algorithms
are more effective than SMO and Naive Bayes.




In summary, we have made the following contributions:
\begin{itemize}
\item We developed \tool, a novel approach that (1) automates the extraction of R-clones from repositories in an accurate way, and (2) recommends clones for refactoring based on a comprehensive set of characteristic features.  
\item We designed 34 features---the largest feature set thus far---to characterize
  the content and evolution of individual clones, as well as the
  spatial, syntactical, and co-evolution relationship among clones.
\item We conducted a systematic evaluation to assess (1) the precision of automatic R-clone extraction, (2) \tool's effectiveness of clone recommendation in comparison with a state-of-the-art approach~\cite{wang2014recommending}, and (3) the tool's sensitivity to the usage of different feature sets and various machine learning algorithms.

\end{itemize}


%% file: approach.tex
\section{Approach}
\label{sec:Approach}
\tool consists of three phases.
In this section, we will first discuss how \tool automatically extracts R-clones and NR-clones from software history (Section~\ref{sec:prepare}). Then we will explain our feature extraction to characterize clone groups (Section~\ref{sec:feature}). Finally, we will expound on the training and testing phases (Section~\ref{sec:train} and~\ref{sec:test}). 

\subsection{Clone Data Preparation}
\label{sec:prepare}
This phase contains two steps. First, \tool conducts clone genealogy analysis to identify clone groups and extract their evolution history (Section~\ref{sec:genealogy}). Second, based on the established clone genealogies, 
\tool implements several heuristics to decide whether a clone group was refactored or not by developers (Section~\ref{sec:label}).

\subsubsection{Clone Genealogy Analysis}
\label{sec:genealogy}
Clone genealogy describes how clones evolve over time. Kim et al. invented an approach to model clone genealogies based on software version history~\cite{kim2005empirical}.
 As the original approach implementation is not available, we reimplemented the approach with some improvement on multi-version
clone detection.
\vspace{-0.5em}
\begin{figure}[!htb]
	\centering
	\includegraphics[width=5.7cm]{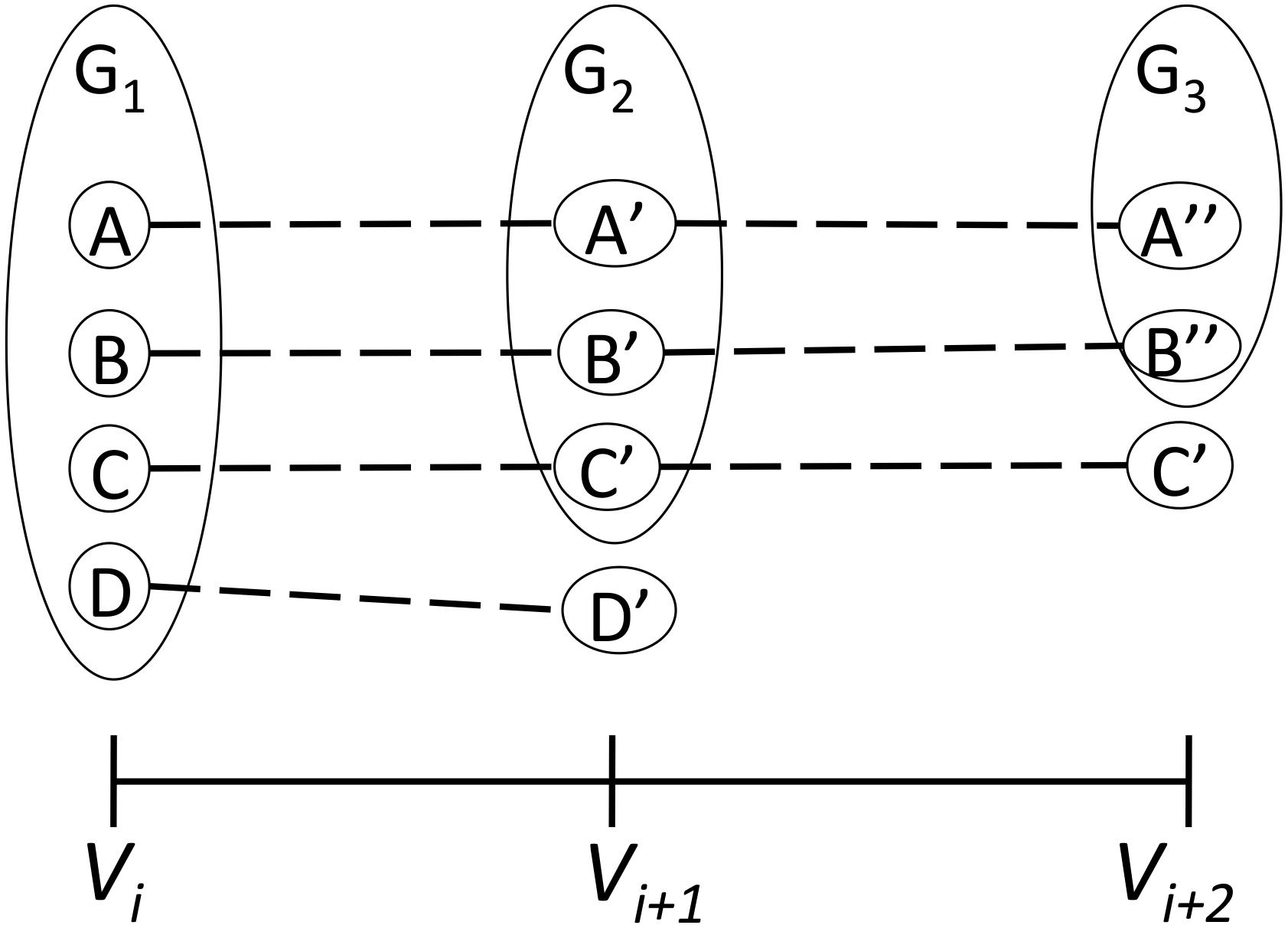}
	\vspace{-0.5em}
	\caption{An exemplar clone lineage}
	\vspace{-5mm}
	\label{fig:genealogy}
\end{figure}

Fig.~\ref{fig:genealogy} illustrates a clone lineage modeled by our analysis. Suppose there are three clone groups detected in a project's different versions: $G_1=\{A, B, C, D\}$, $G_2=\{A', B', C'\}$, and $G_3=\{A'', B''\}$. By comparing clones' string similarity across versions, we learnt that $G_1$ and $G_2$ have three similar but different matching clone members: $(A, A')$, $(B, B')$, and $(C, C')$; while $G_2$ and $G_3$ have two matching members: $(A', A'')$ and $(B', B'')$. 
Therefore, we correlate the three clone groups. 
Furthermore, we infer that $G_1$ evolves to $G_2$ because the majority of their members match. 
$G_2$ evolves to $G_3$, because two members seem to consistently change with the third member ($C'$) unchanged. 
Therefore, there are mainly two parts in our clone genealogy analysis: multi-version clone detection and cross-version clone linking.

\noindent

\textbf{Multi-Version Clone Detection.} 
We utilized SourcererCC~\cite{sajnani2016sourcerercc}, a recent scalable and accurate near-miss clone detector, to find block-level clones. 
As with prior work~\cite{roy2008nicad,sajnani2016sourcerercc}, we configured SourcererCC to report clones with at least 30 tokens or 6 lines of code to avoid trivial code duplication. 
We also set SoucererCC's minimum similarity threshold as 80\% to reduce false alarms.

Since the whole history may contain a huge number of commits, to efficiently establish clone genealogies, we sampled versions in chronological order based on the amount of code revision between versions. 
Specifically, \tool samples the oldest version of a project by default, denoting it as $S_1$. Then \tool enumerates subsequent versions until finding one that has at least 200 changed lines in comparison with $S_1$, denoting it as $S_2$. By selecting $S_2$ as another sample, \tool further traverses the follow-up versions until finding another version 
that is different from $S_2$ by at least 200 lines of code. This process continues until \tool arrives at the latest version.  


\textbf{Cross-Version Clone Linking.}
We used an existing approach~\cite{Wang2014:predictclonecopy} to link the clone groups detected in different sampled versions, in order to restore the clone evolution history. 
This approach matches clones across versions based on the code location and textual similarity. 
For example, suppose that two clones $A$ and $B$ are found in file $F$ of version $V_{i}$; while one clone $A'$ is also found in $F$, but of version $V_{i+1}$. Since $A$ and $A'$ are located in the same file, and if $A$ has more content overlap with $A'$ than $B$ does, we link $A$ with $A'$ across versions, and conclude that $A$ evolves to $A'$. 
With such clone linking, we can further link clone groups across versions based on the membership of linked clones.

\subsubsection{Clone Labeling}
\label{sec:label}
To train a classifier for clone refactoring recommendation, 
we need to label clones either as ``refactored'' or ``not refactored'' by developers. 
Intuitively, if (1) certain clones have duplicated code removed and method invocations added, and (2) the newly invoked method has content similar to the removed code, we infer that developers might have applied ``Extract Method'' refactorings to these clones for duplication removal. 
Based on this intuition, we developed \tool to assess any given clone group $G$ of version $V_i$ with the following three criteria:
\begin{enumerate}
\item[i.] At least two clone instances in $G$ have code size reduced when evolving into clones of version $V_{i+1}$. We refer to this subset of clones as $C=\{c_1, c_2, \ldots\} (C\subset G)$. 
\item[ii.] Within $C$, at least two clone instances (from the same or different source files) add invocations to a method $m()$ during the evolution. 
We refer to the corresponding clone subset as $C'=\{c_1', c_2', \ldots\}$ ($C'\subset C$).
\item[iii.] Within $C'$, at least two clone members have their removed code similar to $m()$'s body.
Here, we set the similarity threshold as 0.4, because the implementation of $m()$ may extract the commonality of multiple clones while parameterizing any differences, and thus become dissimilar to each original clone. 
This threshold setting will be further discussed in Section~\ref{sec:automationPrecision}.

\end{enumerate}
When $G$ satisfies all above criteria, \tool labels it as ``R-clone''; otherwise, it is labeled as ``NR-clone''. 

\subsection{Feature Extraction}
\label{sec:feature}
With the ability to detect so many clones, determining which of them to refactor can be challenging. 
Prior research shows that developers base their refactoring decisions on the potential efforts and benefits of conducting refactorings
~\cite{kim2014empirical, wang2014recommending}. 
Our intuition is that by extracting as many features as possible to reflect the present status and history of programs, we can capture the factors that may affect developers' refactoring decisions, and thus accurately suggest clones for refactoring.
Specifically, we extract five categories of features: two  categories modeling the current version and historic evolution of individual clones, 
and three categories reflecting the location, code difference, and co-change relationship among clone peers of each group. Table~\ref{tab:features} summarizes the 34 features of these 5 categories. 

\input{feature_table}

\subsubsection{Clone Code Features}
\label{sec:content}
We believe that the benefits and efforts of refactoring a clone are relevant to the code size, complexity, and the clone's dependence relationship with surrounding code.
For instance, if a clone is large, complex, and independent of its surrounding context, the clone is likely to be refactored. Therefore, we defined 11 features (F1-F11) to characterize the content of individual clones. 

\subsubsection{Clone History Features}
\label{sec:history}
A project's evolution history may imply its future evolution. 
 Prior work shows that many refactorings are driven by maintenance tasks such as feature extensions or bug fixes~\cite{silva2016we}. 
It is possible that developers refactor a clone because the code has recently been modified repetitively.
To model a project's recent evolution history, we took the latest 1/10 of the project's sampled commits, and extracted 6 features (F12-F17). 

\subsubsection{Relative Location Features among Clones}
\label{sec:location}
Clones located in the same or closely related program components (e.g., packages, files, or classes) can share program context in common, and thus may be easier to refactor for developers.
Consequently, we defined 6 features (F18-F23) to model the relative location information among clones, characterizing the potential refactoring effort.

\subsubsection{Syntactic Difference Features among Clones}
\label{sec:difference}
Intuitively, dissimilar clone instances are usually more difficult to refactor than similar ones, because developers have to resolve the differences before extracting common code as a new method. To estimate the potential refactoring effort, we defined 6 features (F24-F29) to model differences among clones, such as the divergent identifier usage of variables, methods, and types. 

Specifically to extract differences among clones, we used an off-the-shelf tool---MCIDiff~\cite{lin2014detecting}. Given a clone group $G=\{c_1, c_2, \ldots, c_n\}$, MCIDiff tokenizes each clone, leverages the longest common subsequence (LCS) algorithm to match token sequences, and creates a list of matched and differential multisets of tokens. 
Intuitively, if a token commonly exists in all clones, it is put into a \textbf{matched} multiset; otherwise, it is put to a \textbf{differential} multiset.
Furthermore, there is a special type of differential multisets called \textbf{partially same} to record tokens that commonly exist in some ($\ge2$) but not all clones in a group.  
Matched and partially-same multisets may imply zero refactoring effort, because developers do not need to resolve differences before refactoring the tokens' corresponding code.


Meng et al.~showed that variable differences among clones were easier to resolve than method or type differences~\cite{hua2015does}. To more precisely characterize the potential refactoring effort, we also classified identifier tokens into three types: variables, methods, and types.
As MCIDiff does not identify different types of tokens, we extended it to further analyze the syntactic meaning of identifiers.
With more detail, our extension first parsed the Abstract Syntax Tree (AST) of each clone using Eclipse ASTParser~\cite{astparser}. By implementing an ASTVisitor to traverse certain types of ASTNodes, such as 
\codefont{FieldAccess}, \codefont{MethodInvocation}, and \codefont{TypeLiteral}, we determined whether certain token(s) used in these ASTNodes are names of variables, methods, or types. In this way, we further classified identifier tokens and labeled MCIDiff's differential multisets with ``variable'', ``method'', and ``type'' accordingly.

\subsubsection{Co-Change Features among Clones}
\label{sec:co-change}
If multiple clone peers of the same group are frequently changed together, developers may want to refactor the code to reduce or eliminate future repetitive editing effort~\cite{mandal2014automatic}. 
Thus, we defined 5 features (F30-F34) to reflect the co-change relationship among clones.

\subsection{Training}
\label{sec:train}
To train a binary-class classifier for clone recommendation, we first extracted feature vectors for both R-clones and NR-clones in the training set. Then for each feature vector we constructed one data point in the following format: $<feature\_vector, label>$, where $feature\_vector$ consists of \fcount{} features; and $label$ is set to 1 or 0 depending on whether $feature\_vector$ is derived from R-clones or NR-clones.  

We used Weka~\cite{weka}---a software library implementing a collection of machine learning algorithms to train a classifier based on training data. By default, we use AdaBoost~\cite{freund1997decision}, a state-of-the-art machine learning algorithm, because prior research shows that it can be successfully used in many applications~\cite{caruana2006empirical}.
Specifically, we configured AdaBoost to take a weak learning algorithm, decision stump~\cite{iba1992induction}, as its weak learner, and call the weak learner 50 times.


\subsection{Testing}
\label{sec:test}

Given the clone groups in the testing set, \tool{} first extracts features based on the current status and evolution history of these clones. 
Next, \tool{} feeds the features to the trained classifier to decide whether the clone groups should be refactored or not. If a group is predicted to be refactored with over 50\% likelihood, \tool{} recommends the group for refactoring. When multiple groups are recommended, \tool{} further ranks them based on the likelihood outputs by the classifier.

%% file: feature_table.tex
\begin{table*}
\centering
\vspace{-0.5em}
\caption{The 34 features used in \tool}
\label{tab:features}
\vspace{-0.5em}
\begin{tabular}{p{2cm}p{7cm}p{9cm}}
\toprule
\textbf{Feature Category} & \textbf{Features} & \textbf{Rationale}\\
\toprule
\multirow{21}{*}{Clone Code (11)} & F1: Lines of code in a clone. & A larger clone is more likely to be refactored to significantly reduce code size.\\
\cline{2-3}
& F2: Number of tokens. & Similar to F1. \\
\cline{2-3}
& F3: Cyclomatic complexity. & The more complexity a clone has, the harder it is to maintain the clone, and thus the more probably developers will refactor.\\
\cline{2-3}
& F4: Number of field accessed by a clone. & The more fields (e.g., private fields) a clone accesses, the more this clone may depend on its surrounding program context. Consequently, refactoring this clone may require more effort to handle the contextual dependencies. \\
\cline{2-3}
& F5: Percentage of method invocation statements among all statements in a clone. & Similar to F4, this feature also measures the dependence a clone may have on its program context.\\
\cline{2-3}
& F6: Percentage of code lines of a clone out of its container method. & Similar to F4 and F5, this feature indicates the context information of a clone in its surrounding method and how dependent the clone may be on the context.\\
\cline{2-3}
& F7: Percentage of arithmetic statements among all statements in a clone. & Arithmetic operations (e.g., ``+'' and ``/'') may indicate intensive computation, and developers may want to refactor such code to reuse the computation.
\\ \cline{2-3}
& F8: Whether a complete control flow block is contained. & If a clone contains an incomplete control flow, such as an \texttt{else-if} branch, it is unlikely that developers will refactor the code. 
\\ \cline{2-3}
& F9: Whether a clone begins with a control statement. & Beginning with a control statement (e.g., \texttt{for}-loop) may indicate separable program logic, which can be easily extracted as a new method. 
\\ \cline{2-3}
& F10: Whether a clone follows a control statement. & Similar to F9, a control statement may indicate separable logic.
\\ \cline{2-3}
& F11: Whether a clone is test code. & Developers may intentionally create clones in test methods to test similar or relevant functionalities, so they may not want to remove the clones. \\
\hline
\multirow{12}{*}{Clone History (6)}
& F12: Percentage of file-existence commits among all checked historical commits. & If a file is recently created, developers may not refactor any of its code.
\\ \cline{2-3}
& F13: Percentage of file-change commits among all checked commits. & If a file has been changed a lot, developers may want to  refactor the code to facilitate software modification.
\\ \cline{2-3}
& F14: Percentage of directory-change commits among all checked commits. & Similar to F13, this feature also models historical change activities.
\\ \cline{2-3}
& F15: Percentage of recent file-change commits among the most recent 1/4 checked commits. & Compared with F13, this feature focuses on a shorter history, because a file's most recent updates may influence its future more than the old ones.
\\ \cline{2-3}
& F16: Percentage of recent directory-change commits among the most recent 1/4 checked commits. & Compared with F14, this feature also measures how actively a directory was modified, but characterizes a shorter history.
\\ \cline{2-3}
& F17: Percentage of a file's maintainers among all developers. & The more developers modify a file, the more likely that the file is buggy~\cite{Moser:2008:CAE}; and thus, developers may want to refactor such files.\\
\hline
& 
F18: Whether clone peers in a group are in the same directory & It seems easier to refactor clones within the same directory, because clones share the directory as their common context.
\\ \cline{2-3}
& F19: Whether the clone peers are in the same file.& Similar to F18, clones' co-existence in the same file may also indicate less refactoring effort.
\\ \cline{2-3}
Relative Locations& F20: Whether the clone peers are in the same class hierarchy & Similar to F18 and F19, if clones share their accesses to many fields and methods defined in the same hierarchy, such clones may be easier to refactor. 
\\ \cline{2-3}
among Clones (6) & F21: Whether the clone peers are in the same method &Similar to F18-F20, clones within the same method possibly share many local variables, indicating less effort to apply refactorings.
\\ \cline{2-3}
& F22: Distance of method names
measures the minimal Levenshtein distance~\cite{levenshtein1966bcc} (i.e., textual edit distance) among the names of methods containing clone peers. & As developers define method names as functionality descriptors, similar method names usually indicate similar implementation, and thus may imply less refactoring effort.
\\ \cline{2-3}
& F23: Possibility of the clone peers from intentionally copied directories. & 
The more similar two directories' layout are to each other, the more likely that developers intentionally created the cloned directories, and the less likely that clones between these directories are to be refactored. \\

\hline
& F24: Number of instances in a clone group & 
The size of a clone group can affect developers' refactoring decision. For instance, refactoring a larger group may lead to more benefit (e.g., reduce code size), while refactoring a smaller group may require for less effort.  
\\ \cline{2-3}
& F25: Number of total diffs measures the number of differential multisets reported by MCIDiff~\cite{lin2014detecting}. & More differences may demand for more refactoring effort.
\\ \cline{2-3}
Syntactic Differences among Clones (6)& F26: Percentage of partially-same diffs measures among all reported differential multisets, how many multisets are partially same. & Partially-same multisets indicate that some instances in a clone group are identical to each other while other instances are different. Developers may easily refactor the clone subset consisting of identical clones without dealing with the difference(s) in other clone peers. 
\\ \cline{2-3}
& F27: Percentage of variable diffs measures among all differential multisets, how many multisets contain variable identifiers. & If developers resolve each variable diff by renaming the corresponding variables consistently or declaring a parameter in the extracted method, this feature characterizes such effort.
\\ \cline{2-3}
& F28: Percentage of method diffs measures among all differential multisets, how many multisets have method identifiers.& If developers resolve each method diff by creating a lambda expression or constructing template methods, this feature reflects the amount of such effort.
\\ \cline{2-3}
& F29: Percentage of type diffs measures among all differential multisets, what is the percentage of multisets containing type tokens. & Different from how variable and method differences are processed, type differences are possibly resolved by generalizing the types into a common super type. This feature reflects such potential refactoring effort.\\
\hline
& F30: Percentage of all-change commits among all checked commits. & The more often all clones are changed together, the more possibly developers will refactor the code.
\\ \cline{2-3} 
& F31: Percentage of no-change commits among all checked commits. & Similar to F30, this feature also characterizes the evolution similarity among clone peers.
\\ \cline{2-3}
Co-Changes among Clones & F32: Percentage of 1-clone-change commit among all checked commits. & Different from F30, this feature characterize the inconsistency evolution of clones. If clones divergently evolve, they are less likely to be refactored.
\\ \cline{2-3}
(5)& F33: Percentage of 2-clone-change commits among all checked commits. & Similar to F32, this feature also captures the maintenance inconsistency among clones.
\\ \cline{2-3}
& F34: Percentage of 3-clone-change commits among all checked commits. & Similar to F32 and F33.\\
\bottomrule
\end{tabular}
\end{table*}

%% file: results.tex
\section{Experiments}
\label{sec:experiments}
In this section, we first introduce our data set for evaluation (Section~\ref{sec:data}), and then describe our evaluation methodology (Section~\ref{sec:method}). Based on the methodology, we estimated the precision of reporting refactored clone groups (Section~\ref{sec:automationPrecision}). Next, we evaluated the effectiveness of \tool{} and compare it with an existing clone recommendation approach (Section~\ref{sec:compare}) 
Finally, we identified important feature subsets (Section~\ref{sec:FeatureExp}) and suitable machine learning algorithms (Section~\ref{sec:ModelExp}) for clone refactoring recommendation.

\subsection{Data Set Construction}
\label{sec:data}
Our experiments used the repositories of six open source projects: {\sf Axis2}, {\sf Eclipse.jdt.core}, {\sf Elastic Search}, {\sf JFreeChart}, {\sf JRuby}, and {\sf Lucene}.
We chose these projects for two reasons.
First, they have long version histories, and most of these projects were also used by prior research~\cite{tairas2012increasing,wang2014recommending,tsantalis2015assessing,hua2015does,tsantalis2017clone}.
Second, these projects belong to different application domains, which makes our evaluation results representative.
Table~\ref{tab:subjects} presents relevant information of these projects. Column \textbf{Domain} describes each project's domain. \textbf{Study Period} is the evolution period we investigated. \textbf{\# of Commits} is the total number of commits during the studied period. \textbf{\# of Clone Groups} is the total number of clone groups detected in each project's sampled versions.
\textbf{\# of R-clones} means the number of refactored clone groups automatically extracted by \tool. 

To evaluate \tool's effectiveness, we need both R-clones and NR-clones to construct the training and testing data sets. As shown in Table~\ref{tab:subjects}, there are a small portion of detected clone groups reported as R-clones.
To create a balanced data set of positive and negative examples, 
we randomly chose a subset of clone groups from clone genealogies without R-clones. The number of NR-clones in the subset is equal to that of reported R-clones.
Each time when we trained a classifier, we used a portion of this balanced data set as \textbf{\emph{training data}}. For \textbf{\emph{testing data}}, we used the remaining portion of the data set as the testing set, except the R-clones' false positives confirmed by our manual inspection (See Section~\ref{sec:automationPrecision}). We did not manually inspect the NR-clones in the testing set. This is because existing study~\cite{wang2014recommending} has revealed only a very small portion of clone groups (less than 1\%) are actually refactored by developers. Therefore, the portion of false NR-clones (i.e., missing true R-clones) in the testing set cannot exceed 1\%, which would have only marginal effect on our results.

\begin{table*}[!htb]
	\renewcommand{\arraystretch}{1.3}
	\vspace{-0.5em}
	\caption{Subject projects}
	\label{tab:subjects}
	\vspace{-0.5em}
	\centering
	\begin{tabular}{cccccc}
		\toprule
		\textbf{Project} & \textbf{Domain} & \textbf{Study Period} & \textbf{\# of Commits} & \textbf{\# of Clone Groups} & \textbf{\# of R-clones} \\ \toprule
		{Axis2} & core engine for web services &09/02/2004-03/06/2018 &8723 &800950 &43  \\
		{Eclipse.jdt.core} &integrated development environment&06/05/2001-01/22/2018 &22358 &10001106 &106  \\
		{Elastic Search} &search engine &02/08/2010-01/05/2018 &14766 &1856035 &33  \\
		{JFreeChart} & open-source framework for charts &06/29/2007-01/24/2018 &3603 &246098 &59  \\
		{JRuby} & an implementation of the ruby language &09/10/2001-03/04/2018 &24434 &718698 &65  \\
		{Lucene} & information retrieval library &09/11/2001-01/06/2018 &22061 &4774413 &27  \\
		\bottomrule
	\end{tabular}
\end{table*}

\subsection{Evaluation Methodology}
\label{sec:method}
\tool{} automatically extracts clone groups historically refactored (R-clones) from software repositories, and considers the remaining clone groups as not refactored (NR-clones).
Before using the extracted groups to construct the training set,
we first assessed \tool's precision of R-clone extraction. 


Additionally, with the six project's clone data, we conducted within-project and cross-project evaluation. In \emph{\textbf{within-project evaluation}}, we used ten-fold validation to evaluate recommendation effectiveness. In \emph{\textbf{cross-project evaluation}}, we iteratively used five projects' R-clones and NR-clones as training data, and used the remaining project's clone data for testing. In both cases, the clone recommendation effectiveness is evaluated with three metrics: precision, recall, and F-score. 

\emph{\textbf{Precision (P)}} measures among all the recommended clone groups for refactoring, how many groups were actually refactored by developers:

\begin{equation}
P = \frac{ \text{\# of refactored clone groups}}{\text{Total \# of recommended groups}} * 100\%. 
\end{equation}

\emph{\textbf{Recall (R)}} measures among all known refactored clone groups, how many groups are suggested by a refactoring recommendation approach: 
\begin{equation}
R = \frac{\text{\# of refactored clone groups suggested}}{\text{Total \# of known refactored groups}} * 100\%.
\end{equation}

\emph{\textbf{F-score (F)}} is the harmonic mean of precision and recall. It combines $P$ and $R$ to measure the overall accuracy of refactoring recommendation as below:
\begin{equation}
	F = \frac{2 * P* R}{P+ R} * 100\%.
\end{equation}
Precision, recall, and F-score all vary within [0\%, 100\%]. The higher precision values are desirable, because they demonstrate better accuracy of refactoring recommendation. As the within-project and cross-project evaluation were performed with six projects, we obtained six precision, recall, and F-score values respectively, and treated the averages of all six projects as the overall effectiveness measurement.

\subsection{Precision of Reporting R-clones}
\label{sec:automationPrecision}
\tool{} relies on a similarity threshold parameter $l_{th}$ to determine R-clones (see Section~\ref{sec:label}). With the threshold,
 \tool checks whether the implementation of a newly invoked method is sufficiently similar to a removed code snippet.
In our experiment, we changed $l_{th}$ from 0.3 to 0.5 with 0.1 increment, and manually examined the reported R-clones in the six projects to estimate \tool's precision.  

In Table~\ref{tab:precision}, \textbf{\# of Reported} shows the total number of reported R-clones by \tool. \textbf{\#of Confirmed} is the number of reported R-clones manually confirmed. \textbf{Precision} measures \tool's precision. 
As $l_{th}$'s value increases, the total number of reported groups goes down, while the precision goes up. To achieve a good balance between the total number of reported refactored groups and the precision, we set $l_{th}$ to 0.4 by default. This is because when $l_{th}=0.4$, the number of R-clones reported by \tool and the 92\% precision are sufficient for constructing the training set. 


\begin{table}[h]
	\caption{\tool's precision of reporting R-clones}
	\label{tab:precision}
	\begin{tabular}{llll}
		\toprule
		\textbf{Threshold Value} & \textbf{\# of Reported} & \textbf{\# of Confirmed} & \textbf{Precision}\\
		\toprule 
		0.3 & 577 & 460 & 80\% \\
		0.4 & 333 & 307 & 92\% \\
		0.5 & 204 & 193 & 95\% \\
		\bottomrule
	\end{tabular}
\end{table}


\vspace{-0.5em}
\subsection{Effectiveness Comparison of Clone Recommendation Approaches}
\label{sec:compare}
Wang et al.~developed a machine learning-based clone recommendation approach~\cite{wang2014recommending}, which extracted 15 features for cloned snippets and leveraged C4.5 to train and test a classifier. Among the 15 features, 10 features represent the current status of individual clones, 1 feature reflects the evolution history of individual clones, 3 features describe the spatial relationship between clones, and 1 feature reflects any syntactic difference between clones. 
Wang et al.'s feature set is much smaller than our 34-feature set. 
It fully overlaps with our feature category of clone content, and matches three out of our six spatial location-related features. 
Wang et al.'s feature set characterizes individual clones more comprehensively than the relationship between clone peers (11 features vs.~4 features), and emphasizes clones' present versions more than their previous versions (14 features vs.~1 feature). It does not include any feature to model the co-change relationship between clones. 

In this experiment, we compared \tool{} with Wang et al.'s approach to see how \tool{} performs differently.
Because neither the tool nor data of Wang et al.'s paper is accessible even after we contacted the authors, we reimplemented 
Wang et al.'s approach, and used our data set for evaluation. 
As \tool{} is different from Wang et al.'s approach in terms of both the feature set and machine learning (ML) algorithm, we also developed two variants of our approach: \tool{}$_m$ that uses \tool{}'s ML algorithm (AdaBoost) but the feature set in Wang et al.'s work, and \tool{}$_f$ that uses \tool{}'s feature set but the ML algorithm (C4.5) in Wang et al.'s work. 
Next, we evaluated the four clone recommendation approaches with the six projects' data in both the within-project and cross-project settings.
Note that due to different data sets for evaluation, the effectiveness of Wang et al.'s approach in this paper is different from the results reported in their paper.
\begin{table*}[htb] 
	\renewcommand{\arraystretch}{1.3}
	\caption{Effectiveness comparison of different clone recommendation approaches in the within-project setting}  
	\label{tab:withinComparison}
	\vspace{-0.5em}
	\centering
	\scriptsize 
	\begin{tabular}{c|rrr|rrr|rrr|rrr}  
		\toprule   
		& \multicolumn{3}{c|}{\textbf{\tool}} & \multicolumn{3}{c|}{\textbf{\tool{}$_m$}} 
		& \multicolumn{3}{c|}{\textbf{\tool{}$_f$}}& \multicolumn{3}{c}{\textbf{Wang et al.'s Approach}}\\ 
		& \multicolumn{3}{c|}{{\textbf{(\tool{}'s Feature Set+AdaBoost)}}} & \multicolumn{3}{c|}{\textbf{(Wang et al.'s Feature Set+AdaBoost)}} 
		& \multicolumn{3}{c|}{\textbf{(\tool{}'s Feature Set+C4.5)}}& \multicolumn{3}{c}{\textbf{(Wang et al.'s Feature Set+C4.5)}}\\ \cline{2-13}  
		\textbf{Test Project}
		&\textbf{P (\%)}&\textbf{R (\%)}&\textbf{F(\%)}
		&\textbf{P (\%)}&\textbf{R (\%)}&\textbf{F(\%)}
		&\textbf{P (\%)}&\textbf{R (\%)}&\textbf{F(\%)}
		&\textbf{P (\%)}&\textbf{R (\%)}&\textbf{F(\%)}\\ 
		\toprule  
		Axis2 & \textbf{79} & \textbf{86} & \textbf{81}
		& 80 & 70 & 73
		& 74 & 88 & 79 
		& 76 & 64 & 65 \\
		Eclipse.jdt.core & \textbf{85} & \textbf{84}& \textbf{83} 
		& 81 & 79 & 79 
		& 83 & 84 & 82
		& 77 & 79 & 77\\ 
		Elastic Search & \textbf{64} & \textbf{70} & \textbf{66} 
		& 52 & 55 & 50	
		& 69 & 80 & 74 
		& 46 & 45 & 45\\
		JFreeChart & \textbf{91} & \textbf{95} & \textbf{92} 
		& 78 & 78 & 78 
		& 90 & 98 & 94
		& 76 & 83 & 77\\
		JRuby & \textbf{79} & \textbf{87} & \textbf{82} 
		& 66 & 74 & 69	
		& 83 & 81 & 82
		& 69 & 70 & 69\\
		Lucene & \textbf{95} & \textbf{97} & \textbf{95} 
		& 77 & 83 & 79 
		& 93 & 97 & 94 		
		& 93 & 87 & 87\\
		
		\hline  
		\textbf{Average} &  \textbf{82} & \textbf{86} & \textbf{83} 
		& 72 & 73 & 71 
		& 82 & 88 & 84
		& 73 & 71 & 70 \\
		\bottomrule
	\end{tabular}  
\end{table*}

\begin{table*}[htb] 
	\renewcommand{\arraystretch}{1.3}
	\caption{Effectiveness comparison of different clone recommendation approaches in the cross-project setting}  
	\label{tab:crossComparison}
	\vspace{-0.5em}
	\centering
	\scriptsize 
	\begin{tabular}{c|rrr|rrr|rrr|rrr}  
		\toprule   
		& \multicolumn{3}{c|}{\textbf{\tool}} & \multicolumn{3}{c|}{\textbf{\tool{}$_m$}} 
		& \multicolumn{3}{c|}{\textbf{\tool{}$_f$}}& \multicolumn{3}{c}{\textbf{Wang et al.'s Approach}}\\ 
		& \multicolumn{3}{c|}{{\textbf{(\tool{}'s Feature Set+AdaBoost)}}} & \multicolumn{3}{c|}{\textbf{(Wang et al.'s Feature Set+AdaBoost)}} 
		& \multicolumn{3}{c|}{\textbf{(\tool{}'s Feature Set+C4.5)}}& \multicolumn{3}{c}{\textbf{(Wang et al.'s Feature Set+C4.5)}}\\ \cline{2-13}  
		\textbf{Test Project}
		&\textbf{P (\%)}&\textbf{R (\%)}&\textbf{F(\%)}
		&\textbf{P (\%)}&\textbf{R (\%)}&\textbf{F(\%)}
		&\textbf{P (\%)}&\textbf{R (\%)}&\textbf{F(\%)}
		&\textbf{P (\%)}&\textbf{R (\%)}&\textbf{F(\%)}\\ 
		\toprule  
		Axis2 & \textbf{82} & \textbf{84} & \textbf{83} 
		& 62 & 42 & 50 
		& 80 & 53 & 64 
		& 55 & 32 & 40\\
		Eclipse.jdt.core & \textbf{71} & \textbf{90}& \textbf{80} 
		& 57 & 79 & 66 
		& 63 & 86 & 73
		& 67 & 83 & 74\\ 
		Elastic Search & \textbf{70} & \textbf{44} & \textbf{54} 
		& 55 & 50 & 53 
		& 67 & 31 & 43 
		& 48 & 38 & 42\\
		JFreeChart & \textbf{73} & \textbf{86} & \textbf{79} 
		& 39 & 25 & 31 
		& 46 & 20 & 28
		& 33 & 25 & 29\\
		JRuby & \textbf{72} & \textbf{63} & \textbf{67} 
		& 60 & 56 & 58 
		& 75 & 83 & 78
		& 66 & 77 & 71\\
		Lucene & \textbf{93} & \textbf{93} & \textbf{93}
		& 57 & 63 & 60 
		& 100 & 48 & 65 		
		& 48 & 44 & 46 \\
		
		\hline  
		\textbf{Average} & \textbf{77} & \textbf{77} & \textbf{76} 
		& 55 & 53 & 53 
		& 72 & 54 & 58
		& 53 & 50 & 50 \\
		\bottomrule
	\end{tabular}  
\end{table*} 
Table~\ref{tab:withinComparison} presents the four approaches' effectiveness comparison for within-project prediction. On average, \tool{} achieved 82\% precision, 86\% recall, and 83\% F-score, which were significantly higher than the 73\% precision, 71\% recall, and 70\% F-score by Wang et al.'s approach. 
\tool{} improved the overall average F-score of Wang et al.'s approach (70\%) by 19\%.
This implies that compared with Wang et al.'s approach, \tool{} more effectively recommended clones for refactoring. 

To further analyze why \tool{} worked differently from Wang et al.'s approach, we also compared their approach with \tool{}$_m$ and \tool{}$_f$. As shown in Table~\ref{tab:withinComparison}, the F-score of \tool{}$_m$ was a little higher than Wang et al.'s approach while \tool{}$_f$ worked far more effectively than both of them. The comparison indicates that (1) both our feature set and AdaBoost positively contribute to \tool's better effectiveness over Wang et al.'s approach; and (2) our feature set can more significantly improve Wang et al.'s approach than AdaBoost. 


Table~\ref{tab:crossComparison} shows different approaches' evaluation results for cross-project prediction. \tool achieved 77\% precision, 77\% recall, and 76\% F-score on average. In comparison, Wang et al.'s approach acquired 53\% precision, 50\% recall, and 50\% F-score. By replacing their feature set with ours, we significantly increased the precision to 72\%. We also replaced their C4.5 algorithm with the AdaBoost algorithm used in \tool{}, and observed that the F-score increased from 50\% to 53\%. 
By correlating these observations with the above-mentioned observations for the within-project setting, we see that our 34-feature set always works much better than Wang et al.'s 15-feature set. 
Additionally, the replacement of C4.5 with AdaBoost brings small improvement to the effectiveness of Wang et al.'s approach. Therefore, the rationale behind our feature set plays a crucially important role in \tool{}'s significant improvement over the state-of-the-art approach. 

\begin{tcolorbox}
	\textbf{Finding 1:}
	\emph{
		\tool outperforms Wang et al.'s approach, mainly because it leverages 34 features to comprehensively characterize the present and past of code clones.}
\end{tcolorbox}
On average, \tool{} achieved 82\% precision, 86\% recall, and 83\% F-score in the within-project setting, which were higher than the 77\% precision, 77\% recall, and 76\% F-score in the cross-project setting. 
This observation is as expected, because the clones of one project may  share certain project-specific commonality. 
By training and testing a classifier with clone data of the same project, \tool can learn some project-specific information, which allows \tool to effectively suggest clones. 

\tool{} worked better for some projects but worse for the others. 
For instance, \tool{} obtained the highest F-score (95\%) for Lucene and the lowest F-score (66\%) for Elastic Search in the within-project setting. This F-score value range means that for the majority of cases, \tool's recommendation is meaningful to developers. It implies two insights. First, developers' refactoring decisions are generally predictable, which validates our hypothesis that different developers similarly decide what clones to refactor. Second, some developers may refactor clones more frequently than the others and have personal refactoring preferences.
Such personal variance can cause the prediction divergence among different test projects.

\begin{tcolorbox}
	\textbf{Finding 2:}
	\emph{\tool{} suggests refactorings with 66-95\% F-scores within projects and 54-93\% F-scores across projects.
		It indicates that developers' clone refactoring decisions are generally predictable, although the refactoring preference may vary with people and projects.}
\end{tcolorbox}

\begin{figure*}[htb]
	\centering
	\includegraphics[width=18cm]{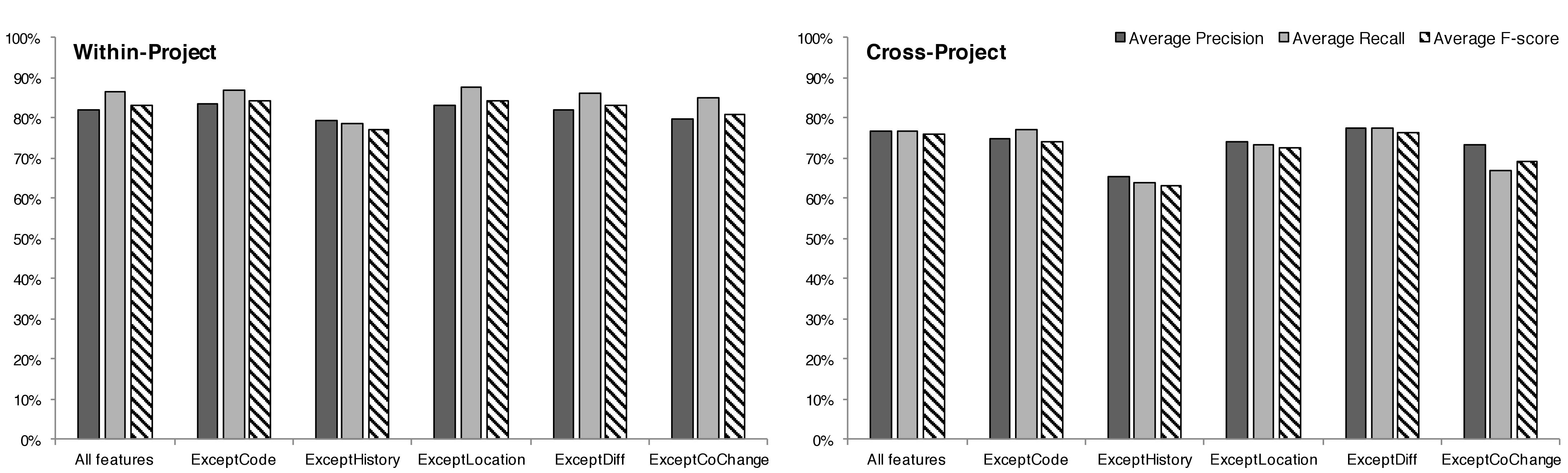}
	\vspace{-0.5em}
	\caption{\tool's average effectiveness when using different feature sets}
	\label{DifferentFeatureSet}
\end{figure*}
\subsection{Important Feature Subset Identification}
\label{sec:FeatureExp}
\tool leveraged 5 categories of features, including 11 features to characterize the content of individual clones, 6 features to reflect individual clones' evolution history, 
6 features to represent the spatial locations of clone peers, 6 features to capture any syntactic difference between clones, and 5 features to reflect the co-evolution relationship between clones. 
To explore how each group of features contribute to \tool's overall effectiveness, we also created five variant feature sets by removing one feature category from the original set at a time. We then executed \tool{} with each variant feature set in both the within-project and cross-project settings (see Fig.~\ref{DifferentFeatureSet}). 

In the within-project setting, \tool's all-feature set achieved 83\% F-score. Interestingly, \textbf{ExceptDiff} obtained the same F-score, which means that the features capturing syntactic differences between clone peers were not unique to characterize R-clones. This category of features was inferable from other features, and thus removing this category did not influence \tool's effectiveness. 
\textbf{ExceptCode} and \textbf{ExceptLocation} obtained the highest F-score---84\%. This may imply that the two feature subsets negatively affected \tool's effectiveness, probably because the code smell and the location information in clones were the least important factors that developers considered when refactoring code.
\textbf{ExceptHistory} obtained the lowest F-score---77\% and \textbf{ExceptCoChange} achieved the second lowest F-score---81\%. This indicates that the evolution history of individual clones and the co-change relationship of clone peers provide good indicators for clone refactorings. 
There seems to be a strong positive correlation between developers' refactoring decisions and clones' evolution history.

In the cross-project setting, \tool's all feature set achieved 76\% F-score. \textbf{ExceptDiff} obtained exactly the same F-score, probably because the features capturing any syntactic differences between clones are inferable from other features. 
All other variant feature sets led to lower F-scores. It means that except the syntactic differences between clones, the other four categories of features positively contribute to \tool's prediction performance. Especially, \textbf{ExceptHistory} obtained an F-score much lower than the scores of other feature sets. This phenomenon coincides with our observations in the within-project setting, implying that the evolution history of code clones 
are always positive indicators of clone refactorings. \textbf{ExceptCoChange} acquired the second lowest F-score, meaning that the co-evolution relationship of clone peers is also important to predict developers' refactoring practice. 
\begin{tcolorbox}
	\textbf{Finding 3:}
	\emph{The clone history features of single clones and co-change features among clone peers were the two most important feature subsets. It seems that developers relied more on clones' past than their present to decide refactorings.}
\end{tcolorbox}


\begin{figure*}[!htb]
	\centering
	\vspace{-0.5em}
	\includegraphics[width=16cm]{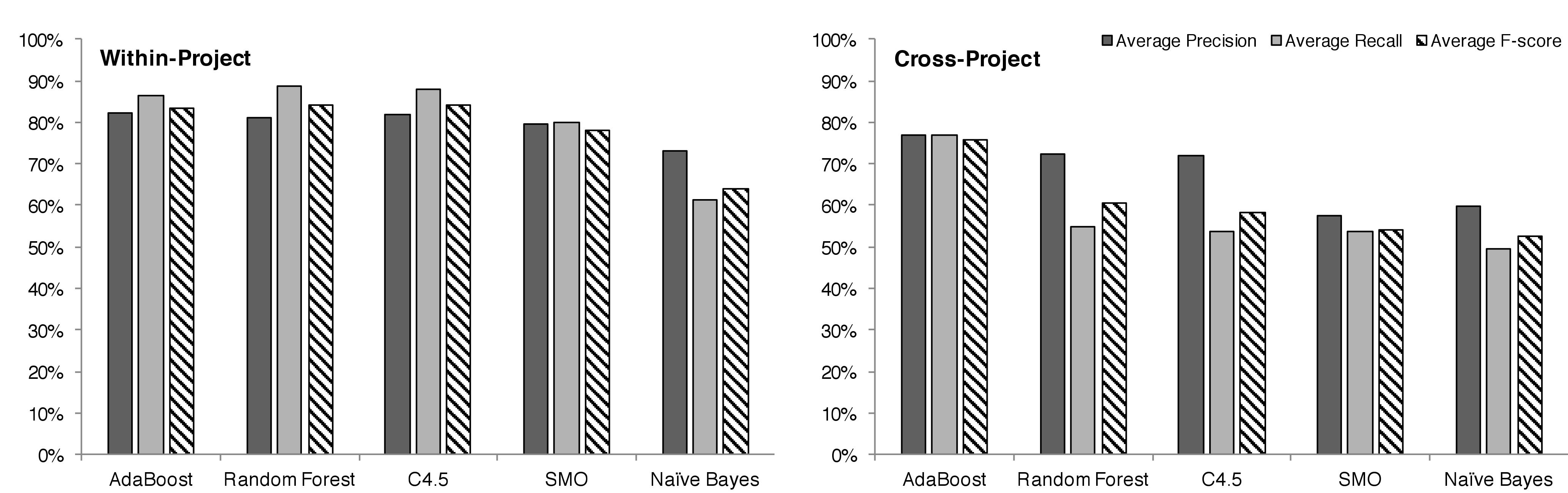}
	\vspace{-0.5em}
	\caption{\tool's average effectiveness when using different ML algorithms}
	\label{fig:DifferentModel}
\end{figure*}

\subsection{Suitable Model Identification}
\label{sec:ModelExp}
To understand which machine learning (ML) algorithms are suitable for \tool{}, in addition to AdaBoost, we also experimented with four other supervised learning algorithms: Random Forest~\cite{Ho:1995}, C4.5~\cite{Quinlan:1986}, SMO~\cite{Platt98:smo}, and Naive Bayes~\cite{Russell:2003}. All these algorithms were executed with their default parameter settings and were evaluated on the six projects for within-project and cross-project predictions. 
Fig.~\ref{fig:DifferentModel} presents our experiment results. The three bars for each ML algorithm in each chart separately illustrate the average precision, recall, and F-score among six projects. 

In the within-project setting, Random Forest and C4.5 achieved the highest F-score: 84\%. AdaBoost obtained the second best F-score: 83\%. 
The F-score of Naive Bayes was the lowest---64\%, and SMO derived the second lowest F-score: 78\%. 
 In the cross-project setting, AdaBoost's F-score (76\%) was much higher than the F-scores of other algorithms. Random Forest obtained the second highest F-score: 61\%. C4.5 acquired the third highest F-score---58\%, while SMO and Naive Bayes had the lowest F-score: 53\%. 
Between the two settings, AdaBoost reliably performed well, while Naive Bayes was always the worst ML algorithm for clone refactoring recommendation. 
AdaBoost, Random Forest and C4.5 are all tree-based ML algorithms, while SMO and Naive Bayes are not. The results in Fig.~\ref{fig:DifferentModel} imply that tree-based ML algorithms are more suitable for clone recommendation than others. 



\begin{tcolorbox}
	\textbf{Finding 4:}
	{AdaBoost reliably suggests clones for refactoring with high accuracy in both within-project and cross-project settings, while Naive Bayes consistently worked poorly. Among the five experimented ML algorithms, tree-based algorithms stably worked better than the other algorithms. }
\end{tcolorbox}




%% file: related_work.tex
\section{Related Work}
The related work includes clone evolution analysis, empirical studies on clone removals, clone recommendation for refactoring, and automatic clone removal refactoring.

\subsection{Clone Evolution Analysis}
Kim et al. initiated clone genealogy analysis and built a clone genealogy extractor~\cite{kim2005empirical}. They observed that extensive refactorings of short-lived clones may not be worthwhile, while some long-lived clones are not easily refactorable. 

Other researchers conducted similar clone analysis and observed controversial or complementary findings~\cite{aversano2007clones,krinke2007study,gode2011frequency,thummalapenta2010empirical,cai2011empirical}. For instance, Aversano et al. found that most co-changed clones were consistently maintained~\cite{aversano2007clones}, while Krinke observed that half of the changes to clones were inconsistent~\cite{krinke2007study}.
G{\"o}de et al. discovered that 88\% of clones were never changed or changed only once during their lifetime, and only 15\% of all changes to clones were unintentionally inconsistent~\cite{gode2011frequency}. 
Thummalapenta et al. automatically classified clone evolution into three patterns: consistent evolution, independent evolution, and late propagation~\cite{thummalapenta2010empirical}.
 Cai et al. analyzed long-lived clones, and reported that the number of developers who modified clones and the time since the last addition or removal of a clone to its group were highly correlated with the survival time of clones~\cite{cai2011empirical}.

These studies showed that not every detected clone should be refactored, which motivated our research that recommends only clones likely to be refactored.

\subsection{Empirical Studies on Clone Removals}
Various studies were conducted to understand how developers removed clones~\cite{gode2010clone}.
Specifically, G{\"o}de observed that developers deliberately removed clones with method extraction refactorings, and the removed clones were mostly co-located in the same file~\cite{gode2010clone}. 
Bazrafshan et al. extended G{\"o}de's study, and found that the accidental removals of code clones occurred slightly more often than deliberate removals~\cite{Bazrafshan2013An}.
Similarly, Choi et al. also observed that most clones were removed with either \emph{Extract Method} or \emph{Replace Method with Method} refactorings. 
Silva et al. automatically detected applied refactorings in open source projects on GitHub, and sent questionnaires to developers for the motivation behind their refactoring practice~\cite{silva2016we}. They found that refactoring activities were mainly driven by changes in the requirements, and \emph{Extract Method} was the most versatile refactoring operation serving 11 different purposes.
None of these studies automatically recommends clones for refactoring or ranks the recommendations.

\subsection{Clone Recommendation for Refactoring}
Based on code clones detected by various techniques~\cite{Kamiya2002, Jiang2007deckard, Krinke2001}, a variety of tools identify or rank refactoring opportunities~\cite{Balazinska1999a, balazinska2000advanced,higo2008metric, Goto2013:rank, Higo2013, Tsantalis2011:ranking}. For instance, Balazinska et al. defined a clone classification scheme based on various types of differences among clones, and automated the classification to help developers assess refactoring opportunities for each clone group~\cite{Balazinska1999a}. 
Balazinska et al. later built another tool that facilitated developers' refactoring decisions by presenting the syntactic differences and common contextual dependencies among clones~\cite{balazinska2000advanced}.
Higo et al.\/ and Goto et al.\/ ranked clones based on the coupling or cohesion metrics~\cite{higo2008metric,Goto2013:rank}.
Tsantalis et al. and Higo et al. further improved the ranking mechanism by prioritizing clones that had been repetitively changed in the past~\cite{Higo2013, Tsantalis2011:ranking}.


Wang et al. extracted 15 features to characterize the content and history of individual clones, and the context and difference relationship among clones~\cite{wang2014recommending}. 
Replacing the feature set with \tool{}'s boost the overall effectiveness of Wang et al.'s approach. More importantly, our research shows for the first time that history-based features are more important than smell-based features (i.e., clone content, location, or differences) when suggesting relevant clones for refactoring.



\subsection{Automatic Clone Removal Refactoring}
A number of techniques automatically remove clones by applying refactorings~\cite{Balazinska1999, Higo2004Refactoring,Juillerat2007, Higo:2012,tairas2012increasing,Tsantalis2013:icsm,bian2013spape,hua2015does,tsantalis2015assessing,tsantalis2017clone}.
These tools mainly extract a method by factorizing the common part and parameterizing any differences among clones.
For example, Tairas et al. built an Eclipse plugin called CeDAR, which unifies clone detection, analysis and refactoring~\cite{tairas2012increasing}. Juillerat et al. and Balazinska et al. defined extra mechanisms to mask any difference between clones~\cite{Balazinska1999, Juillerat2007}.
Several researchers conducted program dependency analysis to further automate \emph{Extract Method} refactorings for near-miss clones---clones that contain divergent program syntactic structures or different statements~\cite{Higo:2012,Tsantalis2013:icsm,Tsantalis2013:icsm,bian2013spape}.
Specifically, Hotta et al.\/ safely moved out the distinct statements standing between duplicated code to uniformly extract common code~\cite{Higo:2012}. Krishnan et al. identified the maximum common subgraph between two clones' Program Dependency Graphs (PDGs) to minimally introduce parameters to the extracted method~\cite{Tsantalis2013:icsm}. Bian et al. created extra control branches to specially handle distinct statements~\cite{bian2013spape}.

Meng et al. detected consistently updated clones in history and automated refactorings to remove duplicated code and to reduce repetitive coding effort~\cite{hua2015does}.
Nguyen et al. leveraged historical software changes to provide API recommendations for developers~\cite{Nguyen2016API}.
Tsantalis et al. automatically detected any differences between clones, and further assessed whether it was safe to parameterize those differences without changing program behaviors~\cite{tsantalis2015assessing}.
Tsantalis et al. later built another tool that utilized lambda expressions to refactor clones~\cite{tsantalis2017clone}.

As mentioned in prior work~\cite{hua2015does}, the above-mentioned research mainly focused on the automated refactoring \emph{feasbility} instead of \emph{desirablity}. 
In contrast, our research investigates the refactoring desirability. By extracting a variety of features that reflect the potential cost and benefits of refactorings to apply, we rely on machine learning to model the implicit complex interaction between features based on clones already refactored or not refactored by developers. In this way, the trained classifier can simulate the human mental process of evaluating refactoring desirability, and further suggest clones that are likely to be refactored by developers. 

%% file: threats.tex
\section{Threats to Validity}
We evaluated our approach using the clone data mined from six mature open source projects that have long version history. Our observations may not generalize well to other projects, but the trained classifier can be applied to other projects' repositories.

We did not manually inspect the non-refactored clones in the testing set, meaning that there may exist some false non-refactored clones. Please note that the false non-refactored clones in the testing set can only be caused by the true refactored clones that were missed by our automatic extraction process. However, according to existing study~\cite{wang2014recommending}, only a very small portion of clone groups (less than 1\%) are actually refactored by developers. Therefore, the portion of false non-refactored clone groups (i.e., missing true refactored clones) in the testing set cannot exceed 1\%, which would have only marginal effect on our results.

By default, \tool{} recommends clones for refactoring when the trained classifier predicts a refactoring likelihood greater than or equal to 50\%. If a given project's version history does not have any commit that changes any clone, it is probable that none of our history-based features are helpful. Consequently, the predicted likelihood can be less than 50\%, and \tool{} does not recommend any clone for refactoring. To at least acquire a ranked list of clones for refactoring in such scenarios, users can tune our likelihood threshold parameter to a smaller value, such as 30\%.

%% file: conclusion.tex
\section{Conclusion}

This paper presents \tool{}, a learning-based approach that suggests clones for \emph{Extract Method} refactoring, no matter whether the clones are located in the same or different files. 

\tool is inspired by prior work, but improves over the state-of-the-art approach by predicting refactorings based on \emph{desirability} instead of \emph{refactorabililty}.
Although it is challenging to mimic developers' mental process of refactoring decisions, \tool (1) conducts static analysis to characterize the present status and past evolution of software, and (2) nicely combines static analysis with machine learning by training a classifier with the analyzed characteristics. 
Our evaluation evidences the effectiveness of our approach. More importantly, we observed that history-based features work more effectively than those features extracted from the current version of software. It indicates a very interesting insight:
when refactoring software, developers consider more on the past evolution than the current version. 
By inspecting the scenarios where \tool cannot predict well, we found that the refactoring decisions seemed to be based on new software requirements or feature additions. In the future, we plan to propose approaches to better handle such scenarios. 
We also plan to conduct user studies with developers to obtain their opinions on the suggested clones for refactoring. 

Compared with prior work, another significant improvement \tool achieves is its automatic extraction of refactored clones from version history. With our tool and data publicly available (\url{https://github.com/soniapku/CREC}), other researchers can access the data, reuse \tool to collect more data as needed, and further investigate refactoring recommendation and automation.